# How DeFi Protocols Choose Oracle Providers: Evidence on Sourcing, Dependence, and Switching Costs

**Giulio Caldarelli**
Department of Management “Valter Cantino”
University of Turin, Italy
giulio.caldarelli@unito.it
**ORCID-ID:** 0000-0002-8922-7871

Abstract

As data is an essential asset for any DeFi application, selecting an oracle is a critical decision for its success. To date, academic research has mainly focused on improving oracle technology and internal economics, while the drivers of oracle choice on the client side remain largely unexplored. This study addresses this gap by gathering insights from leading DeFi protocols, uncovering their rationale for oracle selection and their preferences regarding whether to outsource or internalize data-request mechanisms. Data are collected from founders, C-level executives, and oracle engineers of 32 DeFi protocols, whose combined total value locked (TVL) exceeds 55% of the oracle-using DeFi segment. The study leverages a one-time mixed-method survey, using tailored question paths for in-house versus third-party oracle users. Quantitative answers are summarized, compared across groups, and examined through Spearman rank-order correlations to explore pairwise associations among evaluation dimensions, while open-ended responses are inductively coded into keywords and broader themes to triangulate common selection motives and switching challenges. Insights support the view that protocol choices are tied to technological dependencies, in which the immutability of smart contracts amplifies lock-in, hindering agile switching among data providers. Furthermore, when viable third-party solutions exist, protocols generally prefer to outsource rather than build and maintain internal oracle mechanisms.

## 1. Introduction

“*There is ~40Bn of value secured by these oracles. It’s pretty much the biggest existential risk in all of crypto,*” warns Flare co-founder Hugo Philion (2024). When tens of billions in assets depend on external data feeds, the way DeFi protocols select and govern their oracle providers becomes more than a technical implementation detail. Oracles are a critical infrastructure layer for Web3 applications, and weaknesses in their design, governance, or integration can expose protocols to manipulation, malfunction, and severe economic loss. Industry forensics show that price-oracle manipulation has become one of the most damaging attack vectors in DeFi. Chainalysis estimates that DeFi protocols lost about $403 million to oracle-manipulation attacks in 2022, while later security reports document additional oracle-related losses in 2024 and 2025, bringing cumulative recorded losses to well above $700 million (Chainalysis Team, 2023; Tsentsura, 2025). These figures, however, refer only to misconfiguration, bugs, and manipulation involving price oracles within DeFi protocols. If a broader taxonomy is adopted, in which cross-chain bridges and interoperability layers are treated as oracle-like components relaying state across blockchains, the scale becomes far greater. Chainalysis reports that cross-chain bridge hacks alone accounted for around $2 billion in stolen assets in 2022, while more recent studies and security reports indicate that cumulative bridge-related losses have continued to rise sharply through 2024 and 2025 (Chainalysis Team, 2022, 2025; Hacken, 2025; Wu *et al.*, 2025).

Despite the relevance of these risks, academic and industry research has so far focused primarily on the supply side of the oracle problem, focusing on architectures, reporting mechanisms, incentive design, and cryptographic guarantees (Sztorc, 2015; Murimi and Wang, 2021; Park *et al.*, 2021; Pasdar, Lee and Dong, 2023). While this work is essential, much less is known about the demand side, particularly how DeFi protocols actually choose between alternative oracle solutions, when they decide to build proprietary mechanisms, and how they manage dependence on dominant providers in a highly concentrated market. In other words, the decision problem faced by protocols when selecting an oracle, which we consider at least as important as guaranteeing users' safety in DeFi, has received little attention compared to the technical properties of oracles themselves.

This study addresses that gap by shifting the analytical lens from oracle providers to oracle users. We conceptualize oracle providers as specialized ICT suppliers, and DeFi protocols as buyers engaged in a supplier selection and make-or-buy decision. In this context, we investigate, first, how leading DeFi protocols decide whether to develop proprietary oracle modules or rely on third-party providers. Second, how they evaluate the reliability, governance, and security guarantees of those providers, and third, how they perceive switching costs and lock-in once an oracle is embedded into their protocol design.

To the best of our knowledge, this is the first empirical study to survey DeFi protocols on oracle supplier selection directly and to apply supplier-selection and IT-outsourcing frameworks to this context explicitly. Furthermore, as the area under analysis is vast, consistent with the survey design and thematic structure outlined in Section 3, we organize the study around six thematic areas and corresponding research questions:

**Protocol and sourcing context.** (RQ1) How are oracle sourcing architectures configured among leading DeFi protocols in terms of product type, supported chains, and use of proprietary versus third-party oracle modules?

**Make-or-buy of the oracle module.** (RQ2) What factors drive DeFi protocols to develop proprietary oracle modules instead of relying exclusively on third-party providers?

**Drivers of third-party oracle selection and multi-sourcing.** (RQ3) What criteria do DeFi protocols use to select specific third-party oracle providers, and under which conditions do they adopt multiple providers in parallel?

**Trust, knowledge, and risk allocation in third-party relationships.** (RQ4) How do DeFi protocols understand and evaluate the operation, trustworthiness, and risk-sharing arrangements of third-party oracle providers?

**Switching costs and lock-in.** (RQ5) How do DeFi protocols perceive switching costs and lock-in when considering a change or addition of oracle providers?

**Oracle usage, dependence, and performance.** (RQ6) How do DeFi protocols use oracle services in practice, how dependent are they on them, and how do they assess the performance of proprietary versus third-party solutions?

Empirically, the study draws on survey data from 32 Web3 protocols, whose combined total value locked (TVL) accounted for more than half of the TVL in the economically relevant DeFi universe at the time of data collection. Responses were obtained exclusively from founders, C-level executives, or the engineers directly responsible for oracle integration. Data were collected through a cross-sectional mixed-method questionnaire administered via conditional routing. Protocols relying exclusively on third-party oracles answered a dedicated block on provider selection, whereas protocols using proprietary oracles answered a separate block on in-house development. All respondents completed common items on data needs, perceived quality, satisfaction, and switching costs. Closed-ended items were analyzed through descriptive statistics, structured comparisons across sourcing configurations (proprietary, third-party, and hybrid), and Spearman rank-order correlations to explore pairwise associations among ordinal evaluation variables. Open-ended responses were inductively coded by identifying recurring keywords and grouped into broader themes in order to capture recurrent rationales and constraints. The findings indicate that protocols generally prefer third-party solutions when viable options exist, but in some cases, technological dependencies and specific functional requirements force them to develop and maintain proprietary mechanisms. Even when alternative providers might offer similar or better services, the immutability of smart contracts and the deep integration of oracle logic into protocol architectures can make evaluation and switching costly or practically infeasible. In other cases, protocols simply declare themselves satisfied with their current solution and show little interest in exploring alternatives.

The remainder of the paper is structured as follows. Section 2 reviews the literature on blockchain oracles, supplier selection, IT outsourcing, and switching costs in digital infrastructure markets. Section 3 presents the methodology and survey design. Section 4 reports the empirical findings. Section 5 discusses the results in light of the literature, with particular attention to make-or-buy decisions, multi-sourcing, and vendor lock-in in Web3. Section 6 concludes and outlines directions for future research.

## 2. Literature Background

Blockchain oracles play a critical role in bridging real-world data to blockchain networks, serving as a foundational component in the functionality of smart contracts, particularly in DeFi and other dapps (Heiss, Eberhardt and Tai, 2019; Al-Breiki *et al.*, 2020; Caldarelli, 2023). Their proper functioning prevents failures and hacks that lead to protocol drain or unwanted liquidations. Significant research has been conducted on the technical challenges and solutions related to oracle design, including security issues, data veracity, and decentralization.

### 2.1. Oracle Research and the missing user-side perspective

To date, oracle research has mainly concentrated on the internal properties of oracle systems. Prior studies have explored alternative oracle architectures and their implications for efficiency and reliability (Beniiche, 2020; Mühlberger *et al.*, 2020; Pasdar, Dong and Lee, 2021; Bartholic *et al.*, 2022). In parallel, a line of investigation, guided by the work of Sztorc (2015) and Buterin (2014), leveraged economic theories to find the optimal compromise between decentralization and efficiency, balancing reporter number and consensus mechanisms. Finally, a stream of research focused on the nature of information reported by oracles and investigated the rationale to discover truth with a more philosophical slant. While some studies focused on the relationship between query type and answer quality (Bartholic *et al.*, 2022; Caldarelli and Ornaghi, 2025), others focused on the very concept of ground truth and how to discern it (Egberts, 2017; Damjan, 2018; Frankenreiter, 2019).

Other contributions have examined oracle usage across tokens and chains, as well as the economics and sustainability of oracle projects themselves (Kaleem and Shi, 2021; Liu, Szalachowski and Zhou, 2021; Cong *et al.*, 2025). While this literature is highly valuable, its analytical center remains the oracle protocol itself, so how it is designed, how it performs, and how it secures truthful reporting. By contrast, much less attention has been paid to the user side of the relationship; therefore, how DeFi and other Web3 protocols evaluate alternative oracle providers, what criteria shape adoption decisions, and how those choices affect dependence, switching, and risk allocation over time. This constitutes an important gap, because the robustness of a Web3 protocol depends not only on the technical quality of an oracle per se, but also on how that oracle is selected, integrated, and governed within protocol architecture.

## 2.2 Market selection for technology providers.

Although the blockchain oracle market has not been systematically investigated, DeFiLlama statistics suggest that a small number of large providers account for a substantial share of DeFi TVL. Therefore, we may already hypothesize an oligopoly in the market. If we view the blockchain oracle market as similar to other technology markets, we can draw on numerous academic studies that highlight the conditions leading to market concentration. A key concept to understand the dominance of a few technology providers in technology markets is the network effect. A network effect exists when the value of a product increases as more people use it (Jullien, Pavan and Rysman, 2021). Network effects encourage the standardization of a single technology or the monopolization of a single network, a phenomenon also known as market tipping.

These network effects often create highly concentrated markets, sometimes characterized by a natural monopoly, a situation in which one firm can serve the entire market at a lower cost than multiple firms due to economies of scale. As Katz and Shapiro (1994) explain, in cases in which fixed costs are high and marginal costs tend to zero, markets tend to favor a single provider.

Although initially convenient for users, this also creates negative conditions. As Salop (2021) explains, in the absence of regulation, the same forces that make digital products efficient can also create major barriers to entry, yielding a quasi-monopolistic market structure. Liebowitz and Margolis (1994) also observe that, if a supplier is selected due to network effects, this raises questions about whether it really offers the best technical solution among the available alternatives. In our case, it is interesting to understand if there is a monopolist among oracle providers, what factors contributed to this dominant position, and how it is perceived by its users. It is also crucial to know if it is preferred over the available alternatives.

It is important to restate that these monopolies are natural in the sense that they are not the outcome of malevolent collusion between capitalist forces or regulatory bodies but a direct consequence of market forces. Users basically gravitate to the largest and most valuable network, and the winner can capture most of the market simply by being ahead. Well-known examples of these conditions are Google, Facebook, and Amazon in their respective domains.

As these conditions are known and their outcomes are predictable, they change the rules of competition. Labeled in literature as *“bidding for the future”,* an actor may therefore subsidize its service, with the aim of benefiting from a monopoly later on (Goeree, 2003). Basically, the competition shifts from “in” the market to “for” the market.

Another important factor to consider in market selection is the switching costs and lock-in. When technologies are incompatible with each other, users who commit to a provider or a standard face significant costs (financial, technical, or convenience costs) to switch to an alternative. Switching costs and network effects can “lock” the market, binding users or even the whole market to an early choice. This lock-in grants the incumbent significant ex post market power over its user base (Farrell, 2004). Subsequent competitors face not only the daunting task of offering a superior product but also the challenge of overcoming user expectations and network lock-in. In our case, for oracles, we have to investigate if there is actually a real barrier that prevents others from entering the market after a natural monopoly is established. Even if the dominant provider currently offers the best service, this does not guarantee that its performance will remain optimal over time. Technological dependencies may therefore impede superior future providers from entering the market, obliging users to a lower level of service.

A historical example is the QWERTY keyboard layout, for which David (1985) argued that it became entrenched due to a historical accident and switching costs, even though the Dvorak layout was more efficient. Although it's still to be demonstrated which of the two technologies is actually better, it's arguable that market selection can sometimes favor a technology that isn't optimal. Once a standard achieves critical mass, the benefits of joining the majority can outweigh the individual benefits of a superior but less adopted alternative.

It's also important to note that not all networked markets inevitably end up with a single standard or monopoly provider. Research also identifies conditions where multiple standards may coexist. If network effects diminish after a certain point or if consumer preferences are heterogeneous, the market can sustain more than one viable network (Page and Lopatka, 2000). A classic example that remains valid today is the coexistence of Windows and macOS as separate standards. Although Windows is the dominant mass-market standard, macOS is the standard for a specific product niche due to its differentiation. The same can be said for the duopoly of Android and iOS. These examples show that multi-homing users can mitigate the winner-takes-all effect. If some users strongly prefer an alternative or if platforms are not entirely compatible, an oligopoly can persist rather than a full monopoly. In the blockchain context, we know that different networks and standards coexist, such as Bitcoin, Ethereum, and so on. These coexisting standards may also support the coexistence of different oracle standards, de facto facilitating an oligopoly. Investigating different network dependencies may also shed light on this aspect.

Although the literature on technology suppliers rarely considers this aspect, Web3 protocol developers may also be capable of building their own data modules, it is also interesting to understand the drivers behind in-house development of the oracle module or its externalization to a third party.

The supplier-selection literature has employed different methods to investigate these markets, and the following section provides insights into how we decided to combine and adapt prior rationales to our case.

## 3. Data and Method

As this is an exploratory study of market selection in DeFi, this section explains how the data were collected, the barriers encountered in this process, and how they were overcome. Because Web3 protocols differ from conventional firms in several respects, the present section also explains how the methodology was tailored to this specific study.

### 3.1. Research design and unit of analysis.

This study adopts an exploratory mixed-method, cross-sectional survey design to investigate how DeFi protocols select oracle solutions and how they perceive related sourcing risks and service performance. The unit of analysis is the DeFi protocol, as we submitted one survey per protocol and targeted respondents among individuals directly responsible for oracle decisions and/or implementation, including founders, C-level executives, and delegated oracle engineers.

To ensure the investigation is not descriptive-only, the questionnaire is structured around established constructs from supplier selection and IT outsourcing research, treating oracles as service suppliers (Dickson, 1966; Weber, Current and Benton, 1991; de Boer, Labro and Morlacchi, 2001). Thematic blocks map each construct into specific variables and question types. Table 1 outlines the thematic blocks and literature anchors of our survey design.

**Table 1**. Survey Structure, Variables, and Literature Anchors

| Thematic Areas | Main Variables / Question types | Literature anchors |
| --- | --- | --- |
| Protocol and Sourcing Context | Respondent role; Protocol product type; Supported Chains; Oracle data module structure. | (Dickson, 1966; Weber, Current and Benton, 1991; de Boer, Labro and Morlacchi, 2001) |
| Make-or-Buy of the oracle module | Proprietary or third-party module; Reasons for building an in-house Oracle (e.g., | (Aubert, Rivard and Patry, 2004; Opara-Martins, Sahandi and Tian, 2016) |

| | reputation, security); barriers for an in-house Oracle (open-ended). | |
|---|---|---|
| Drivers of third-party oracle selection and multi-sourcing | Reasons for choosing specific third-party (e.g., innovation, technical features); Reasons for relying on multiple third-parties (e.g., multi-chain support, redundancy for security); Questions on pricing model. | (Weber, Current and Benton, 1991; Handley, Skowronski and Thakar, 2022) |
| Trust, knowledge, and risk allocation in third-party relationships | Binary and Likert items on whether respondents know how the oracle works, who the data reporters are, perceived trustworthiness of reporters, and compensation in case of manipulation/misreporting. | (Kim and Chung, 2003; Hanafizadeh and Zare Ravasan, 2018) |
| Switching costs and lock-in | Scenario question on willingness to switch provider if another offers the same service at a lower price; Likert item on perceived difficulty of changing or adding an oracle; Open-ended questions probing technical and governance barriers to switching. | (Aubert, Rivard and Patry, 2004; Martin, 2012; Schneider and Sunyaev, 2014) |
| Oracle usage, dependence, and performance | Frequency of data requests; type of data requested; Likert items on quality of service (e.g., timeliness, completeness); Likert items on relative ratings for decentralization, security, cost, legal, etc. | (de Boer, Labro and Morlacchi, 2001; Kim and Chung, 2003; Wu and Weng, 2010) |

The survey is available in its entirety as supplementary material, but the complete dataset is kept private due to respondents' requests. The following subsection explains how the data for this research were collected and how we selected the entities and people to fill out the survey.

### 3.2. Data collection

Identifying a relevant and economically meaningful population of DeFi protocols for our study is a non-trivial task. Many listed protocols, in fact, do not require external oracles to operate. For instance, automated market makers such as Uniswap derive asset prices directly from on-chain liquidity pools and can themselves function as decentralized oracles for other protocols. Likewise, liquid staking or restaking services such as Lido or EigenLayer rely on built-in oracle mechanisms and thus cannot meaningfully choose between proprietary and third-party oracle solutions. Therefore, using the DeFiLlama.com oracle section, we restricted the sampling frame to protocols that can, in principle, choose between proprietary and third-party oracle solutions.

On 7 February 2025 (research initiation date), the oracle-related DeFi segment accounted for USD 58.3B out of USD 114.41B total DeFi TVL on DeFiLlama. Protocols in this universe vary widely in size, as some manage several billions, while others have a few thousand or no capital at all. Because Web3 protocols are not conventional companies and their smart contracts remain immutably deployed on the blockchain, they do not "fail" in the traditional sense and can remain "listed" on indexing websites even if abandoned due to hacks, rug pulls, or developer exit. To improve comparability and analytical relevance, we therefore retained only the protocols with at least USD 10M TVL, obtaining 73 protocols (combined USD 49.8B TVL) as the target population. Although $10M is an arbitrary value, we believe it's an appropriate value given that the mean and median are unreliable due to the massive presence of protocols with $0 (or nearly $0) TVL. All 73 protocols were invited, although five declined due to legal/privacy concerns, resulting in an accessible universe of 68 protocols. After multiple contact attempts and follow-up requests for incomplete surveys, we obtained 32 completed surveys, representing USD 32.3B TVL, corresponding to 55.4% of the TVL in the DeFined population as of 7 February 2025.

There are also other factors to consider to better evaluate the value of the collected data. Because Web3 protocols often lack formal organizational structures, official contacts, or identifiable procurement functions, data collection required protocol-by-protocol outreach, including online communities and in-person industry events. In addition, we sought responses from the highest-level representatives (founders/CEO/CTO) or, where delegated, from engineers explicitly responsible for oracle integration. Moreover, this research was not financially or operationally supported by any third-party research company or foundation; therefore, no help was received in collecting the data. Due to these barriers and constraints, data collection required sustained time and effort and required almost a year to be finalized.

Finally, it is useful to clarify the rationale for the study design, to enable a fair comparison with supplier-selection research in more mature settings. First, because oracle sourcing is an emerging and under-researched decision problem in Web3, we adopt an exploratory, key-informant survey approach aligned with foundational supplier-selection research (Dickson, 1966; Choi and Hartley, 1996). This design is intended as a market-mapping and construct-discovery step. It establishes initial evidence on decision-makers' selection criteria, sourcing configurations (outsourcing vs in-house), and perceived switching challenges, and it helps identify which variables are observable and which mechanisms require deeper causal or configurational analysis in subsequent studies. In other words, before applying alternative empirical designs, it is necessary to validate the feasibility of data collection, the accessibility of respondents, and the interpretability of constructs in this novel context.

Second, the achieved sample size should be interpreted in light of the bounded size of the eligible population. Many classic survey studies in supplier selection and IT outsourcing operate within substantially larger sampling frames, often involving hundreds of firms or respondents (Grover, Cheon and Teng, 1996), which can yield large achieved samples even when response rates are modest. In contrast, the oracle-relevant population in our setting is materially smaller (73 eligible protocols in the DeFined universe). Consequently, even with an acceptable response rate (43%), the absolute number of observations remains limited. This constraint is inherent to the setting and should be considered when comparing this study to survey evidence from more established industries.

### 3.3. Survey instruments and measures

Given the exploratory purpose, the bounded population size, and the predominance of categorical and ordinal measures, we combined descriptive and small-sample nonparametric analyses. Categorical variables (e.g., sourcing configuration, pricing model, knowledge of reporters) are summarized as counts and percentages. Likert-type items are reported using distributions and central tendency measures. Service performance was captured through four 1–5 agreement items (timeliness, completeness, correctness, overall satisfaction; 1 = strongly disagree, 5 = strongly agree), which were aggregated into a composite Satisfaction Index (mean of items) and assessed for internal consistency using Cronbach's alpha. For ordinal outcomes compared across sourcing configurations (proprietary vs third-party vs hybrid), we use Kruskal–Wallis tests with pairwise Mann–Whitney U follow-ups and Holm correction. To explore pairwise associations among ordinal variables, we computed Spearman rank-order correlation coefficients (ρ) across key evaluation items. Because third-party and proprietary oracle evaluations are answered by different subsets of the sample and refer to different objects of assessment, separate correlation matrices were computed for each. Spearman's ρ was selected because the variables are predominantly ordinal (Likert-type 1–5 scales) and the sample size does not support distributional assumptions required by parametric alternatives. The resulting correlation matrix is reported and selectively discussed, where pairwise associations reinforce or qualify the descriptive and comparative findings. For within-protocol paired comparisons available in hybrid protocols (e.g., third-party vs proprietary evaluations), we use Wilcoxon signed-rank tests. The "relative comparison" follow-ups (less / about the same / more) are treated as ordinal categorical variables and analyzed descriptively through their category distributions and, where informative, through contingency-table comparisons. Open-ended responses were analyzed through an inductive thematic coding procedure. Initial keywords were identified from each response and then consolidated into broader themes. Theme frequencies were reported to support the interpretation of quantitative patterns, while non-informative answers were excluded from thematic counts. Table 2 provides an illustrative extract of the coding procedure, showing how individual responses were translated into initial keywords and assigned themes.

Table 2. Coding Procedure extract

| R_Id | Q_id | Q_T | Response_excerpt | Keywords | Assigned_themes |
|---|---|---|---|---|---|
| P2 | Q7 | PMA | "…certain tokens that don't meet the necessary liquidity criteria" | push feed; illiquid tokens; asset coverage | Asset coverage |

| P4 | Q7 | PMA | “…validate prices against an on-chain time-weighted average price … fallback to on-chain prices if the reporter key is compromised…” | permissionless feed; TWAP validation; multiple reporters; fallback mechanism | Security and manipulation resistance; Custom pricing methodology |
|---|---|---|---|---|---|
| P4 | Q8 | PMC | “Ensuring reliability and security … tune the TWAP window and acceptance threshold … perform extensive audits…” | reliability; security; TWAP tuning; audits | Reliability and security robustness; Parameter tuning and methodology |
| P4 | Q46 | SC | “…rewriting key price-oracle contracts … passing a governance proposal … integrated across multiple networks…” | contract rewrite; feed reconfiguration; governance proposal; multi-network integration; audits | Technical integration and code changes; Governance and organizational approval; Security and risk aversion |
| P12 | Q46 | SC | “…changing them requires withdrawing all TVL and going through significant changes…” | oracle dependency; infrastructure reliance; TVL withdrawal; switching disruption | Architectural lock-in / immutability |
| P16 | Q25 | TPMB | “…there is always some variation on the price before an update occurs” | dependency; update lag; price variation | Price accuracy and liquidation risk; Dependency on provider |
| P22 | Q25 | TPMB | “There really are no truly 100% decentralized oracles on the market … oracle teams retain some form of … control” | decentralization; admin control | Centralization and admin control |

R_id = Respondents ID, Q_id = Question ID, Q_T = Question Theme, PMA = Proprietary_Main_Advantage, PMC = Proprietary_Main_Challenge, SC = Switching_Costs, TPMB = Third-Party_Main_Barrier

## 4. Findings

The findings are organized by thematic area, consistent with the structure outlined earlier. First, we present data on demographics and supplier composition; then, we explore the dynamics of make-or-buy decisions, outsourcing risks, switching costs, and lock-in. Finally, we provide an overview of consumer perceptions of the service offered and a comparison between suppliers. To improve readability, the main text reports the core results, while complete inferential test tables and supplementary tables are provided in the Appendix.

### 4.1 Demographics of the sample

Although founders and C-level executives (CEO/CTO) were the primary targets of the survey, in several cases, they reported not being fully informed about how the oracle worked and redirected the questionnaire to specialized oracle engineers. In at least seven protocols, we identified dedicated “oracle experts” responsible for this component (Table 3). This suggests that, despite the high level of technical sophistication of DeFi founders, oracle management is often delegated to specialized personnel.

**Table 3**. Respondent Role in the DeFi protocol

| Role | n | % |
|---|---|---|
| Founder | 8 | 25 |
| Engineer | 7 | 21.9 |
| CEO | 7 | 21.9 |
| Executive team | 5 | 15.6 |

| CTO | 5 | 15.6 |
|---|---|---|

Regarding oracle selection, as shown in Table 4, in many protocols, this decision is handled by the founders or CTO, while the rest opt for collegial decision-making, with a small part organized as a DAO with decentralized decision-making.

**Table 4**. Oracle selection process.

| Answer choice | n | % |
|---|---|---|
| Made by Founders/CTO: The decision is made solely by the company's founders or Chief Technology Officer. | 13 | 40.6 |
| Outcome of Board Discussion: The decision is made collaboratively during board meetings with input from various stakeholders. | 4 | 12.5 |
| Votes by Members as a DAO: The decision is made democratically with votes from members of a Decentralized Autonomous Organization (DAO). | 5 | 15.6 |
| Consultation with External Advisors: The decision is influenced by recommendations from external consultants or industry experts. | 4 | 12.5 |
| Executive Management Team Decision: The decision is made by senior management team members, separate from the founders or board. | 6 | 18.8 |

Figure 1a summarizes the products offered by the surveyed protocols, while Figure 1b lists the chains on which they operate. Most protocols in the sample belong to the DeFi lending sector, with a substantial share also active in asset management. Protocols primarily operate on EVM-compatible chains, especially Ethereum and its Layer-2 networks, as well as other alternatives such as Solana and Avalanche.

**Figure 1a**. Product/s offered by the Web3 protocol

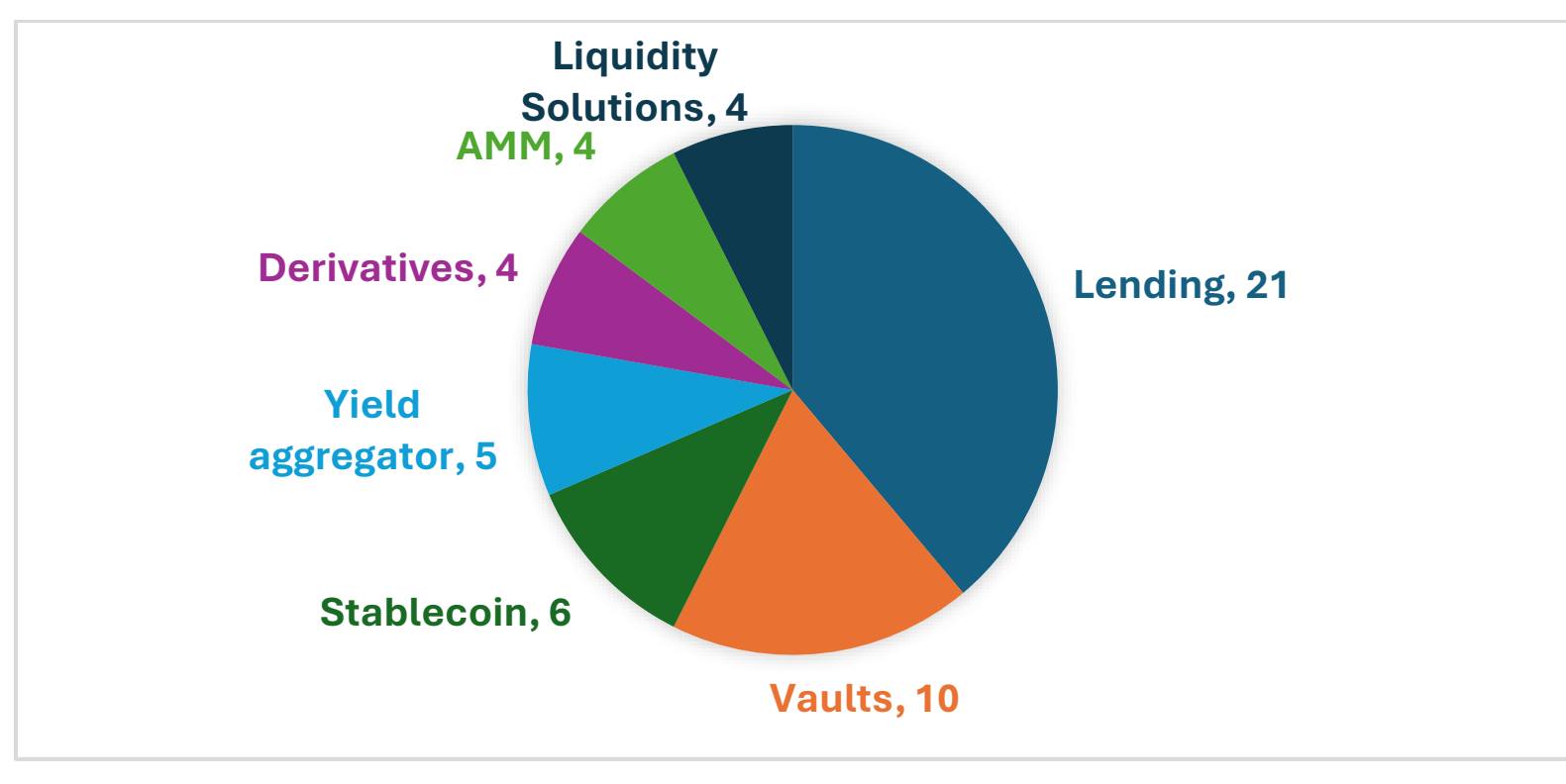

**Figure 1b**. Chains on which your protocol operates.

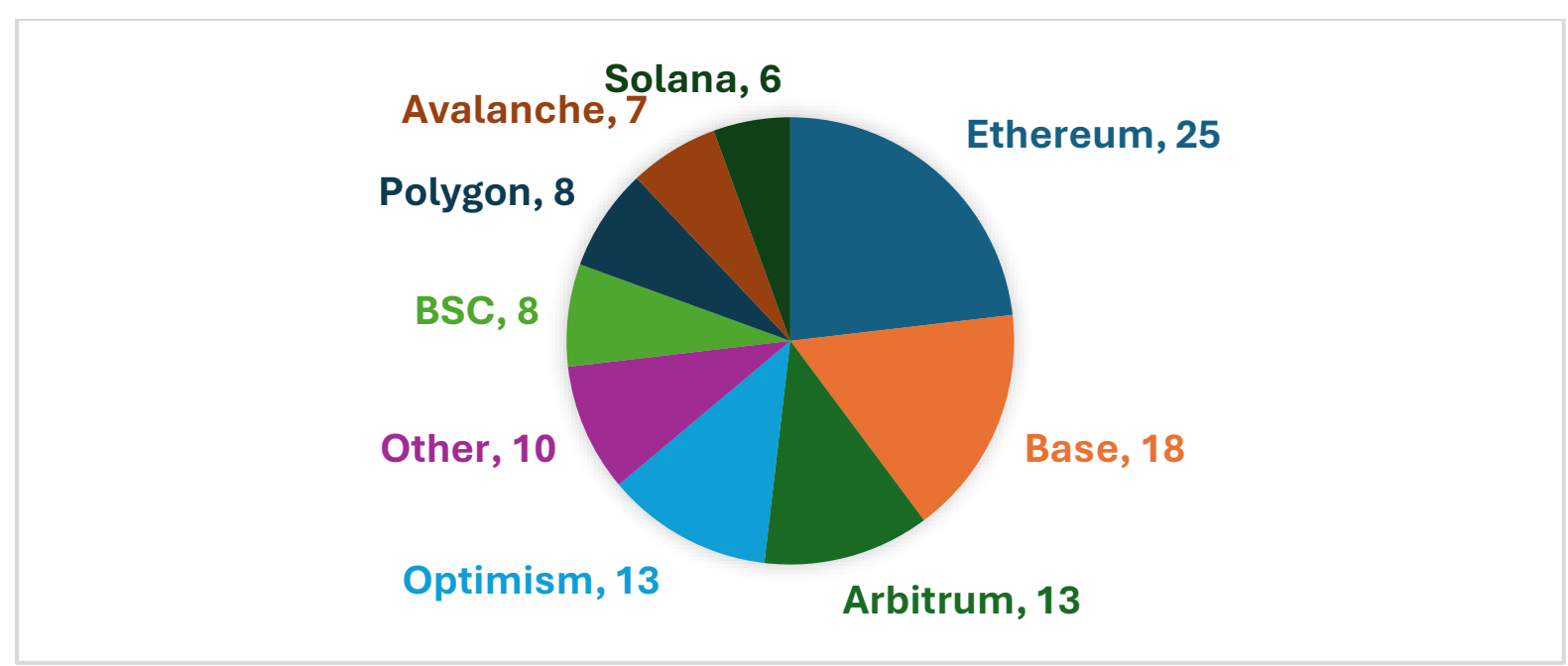

Table 5 shows the distribution of oracle structures adopted by the protocols. As illustrated, DeFi protocols frequently rely on multiple oracle providers. Almost half of the sample uses more than one third-party oracle, whereas roughly one third combines a proprietary oracle module with multiple external providers. This already hints at a complex sourcing landscape characterized by both in-house development and multi-sourcing strategies.

**Table 5**. Web3 protocol oracle structure.

| OracleStructure | n | % |
|---|---|---|
| Third-party (multiple) | 14 | 43.75 |
| Proprietary + Third-party (multiple) | 10 | 31.25 |
| One Third-party | 5 | 15.63 |
| Proprietary + One Third-party | 2 | 6.25 |
| Proprietary (only) | 1 | 3.13 |

Figure 2 finally presents the oracle market composition. The data show a dominant provider serving almost the entire sample, while a small number of alternative providers each serve fewer than half of the respondents. The leading provider is almost always present in the oracle configuration, even when other providers are simultaneously contracted, and regardless of the network on which the protocol operates. This suggests that the dominant provider can support multiple technical standards across different networks. At the same time, alternative providers are added when additional, specific, or niche requirements must be addressed. These aspects are further clarified by the qualitative evidence discussed in the following subsections, which also examine the interplay between third-party and proprietary oracles.

**Figure 2**. Market Composition

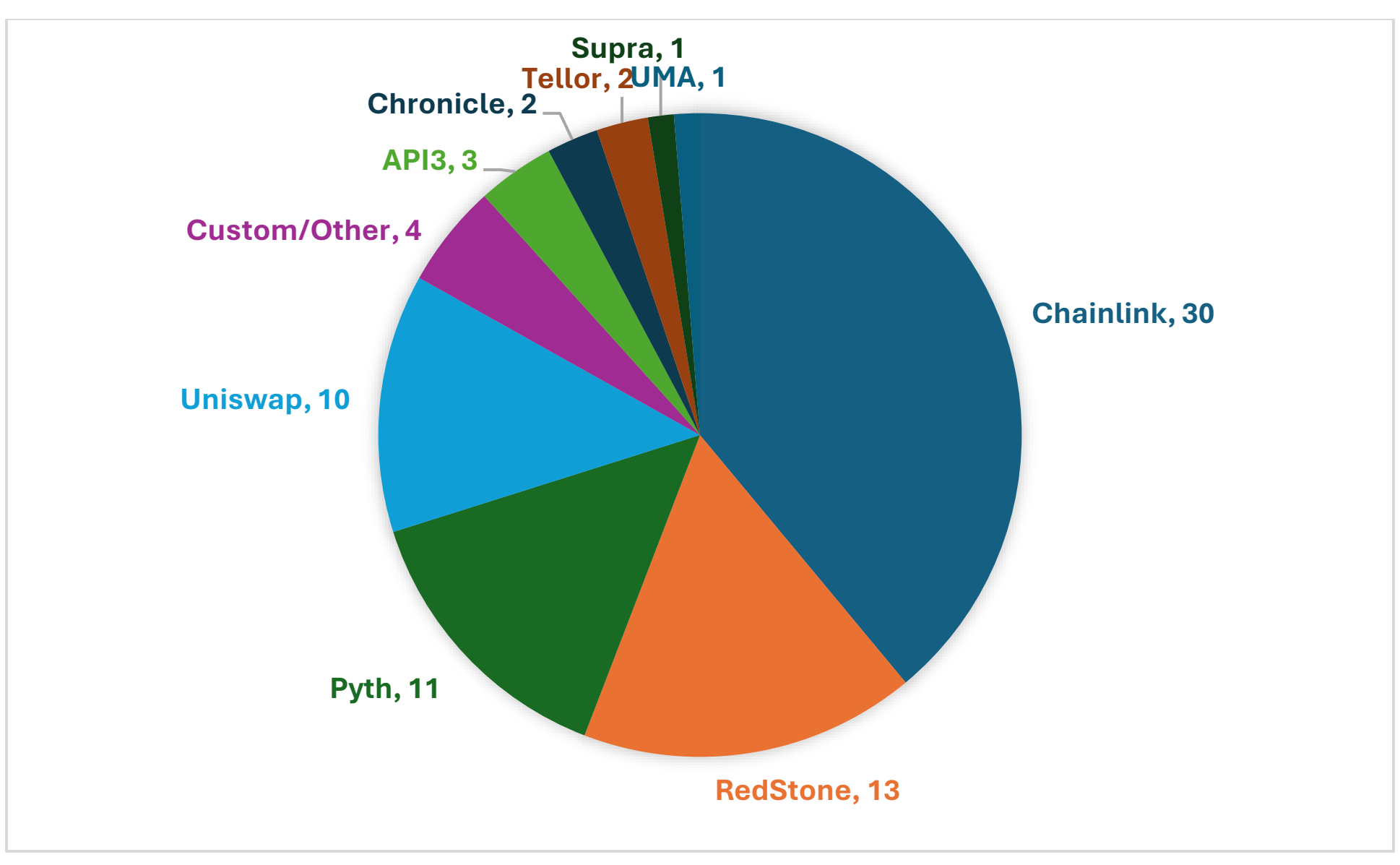

### 4.2 Proprietary vs. Third-Party oracle solutions

This section first analyzes the motives for building in-house oracles, then delves into the advantages and challenges of leveraging third-party solutions.

#### 4.2.1 In-house oracle solutions: Drivers and barriers

As shown in Table 6, protocols primarily develop their own data-collection mechanisms to achieve greater customization, followed by control and integration motives. In other words, the decision to build in-house is mainly driven by technical requirements, although a subset of respondents also mentions security as a decisive factor. By contrast, costs, independence from external providers, and regulation do not emerge as major concerns when opting for internal solutions.

**Table 6**. Reasons for building a proprietary oracle

| Answer choice | n | % |
|---|---|---|
| Customization: Ability to tailor the oracle to specific business needs and requirements. | 12 | 92.31 |
| Control: Greater control over the data and its sources, ensuring data privacy and security. | 7 | 53.85 |
| Integration: Easier integration with existing internal systems and processes. | 6 | 46.15 |
| Data Security: Enhanced security measures that are not available with third-party oracles. | 5 | 38.46 |
| Cost Efficiency: Long-term cost savings compared to using third-party services. | 2 | 15.38 |
| Independence: Reducing dependency on external providers and avoiding potential disruptions. | 2 | 15.38 |
| Regulatory Compliance: Meeting specific regulatory requirements that third-party oracles cannot comply with. | 0 | 0 |

However, generic references to “technical reasons” are insufficient to understand why protocols feel compelled to develop their own oracles and what exactly third-party providers fail to deliver. For this reason, open-ended questions were used to elicit more detailed explanations, and coding was leveraged to triangulate closed question selection.

As coding shows (Table S3), a recurrent theme concerns the presence of assets that do not meet liquidity requirements for inclusion in mainstream oracle price feeds ("Asset Coverage" with 6 keyword occurrences). These assets tend to be thinly traded and therefore excluded by major oracle providers. Another common issue is the absence of USD-denominated prices for specific tokens ("Custom Pricing" with 4 keyword occurrences). As one respondent notes, "*not all tokens have price feeds in USD, so customized solutions were to use existing feeds to calculate USD price*". In such cases, protocols build proprietary mechanisms to derive USD prices from available feeds.

Several respondents also highlight the need for enhanced security and robustness against price errors ("Security and manipulation resistance with 4 keyword occurrences), but most emphasize the necessity of implementing specific price-discovery features that third-party oracles did not support at the time of protocol launch. One CTO explains:

"*We needed a permissionless feed that could validate prices against an on-chain time-weighted average price……and reject values that deviate beyond a set threshold. It also allows adding multiple reporters and has a fallback to on-chain prices if the reporter key is compromised; features unavailable with third-party oracles.*"

In other cases, protocols initially developed features such as exponential weighted moving averages before these were later integrated by third-party providers (e.g., Pyth). Even where such features are now available externally, respondents report that switching from a well-tested internal solution is no longer considered convenient.

The historical dimension is also relevant. A founder of a well-known protocol notes:

"*When our protocol was launched, there were no third-party oracle products available. But even now, we view dependence on real-time market price feeds as a security and structural liability that should be avoided. We use a 24-hour volume-weighted average price, which is not supported by any third-party oracle.*"

Thus, for some early protocols, the in-house solution is both a legacy of a period when third-party options were unavailable and a deliberate design choice to avoid perceived risks associated with real-time feeds.

Open-ended questions also explored the barriers associated with building proprietary oracles. While some respondents simply cite "time and costs", others provide more detailed accounts of technical challenges. One representative explains:

"*We had to tune the TWAP window and acceptance threshold for each market, migrate the anchor from Uniswap v2 to v3 as liquidity moved, and perform extensive audits to prevent* *manipulation. Managing governance votes and multisig updates for upgrades adds additional complexity.*"

These accounts highlight the ongoing maintenance burden of proprietary oracles, including parameter tuning, source migration, security audits, and governance overhead, themes that recurred at least three times during the answer coding.

Interestingly, not all respondents perceive in-house development as particularly demanding. One representative from a pioneering DeFi protocol states:

"*It's very easy to build the oracle, and there's no magic involved. We view the job of the oracle as purely a data pipe, and not a system that should own 'market trustworthiness' or anything like that. The challenge is building a protocol that's robust against oracle failures and/or noisy data.*"

This perspective contrasts sharply with the idea that the oracle itself should be the primary locus of trust. Instead, it shifts responsibility to the protocol design, which must remain robust to potential oracle failures or noisy inputs. This view aligns with the finding (Table 7) that none of the protocols have developed internal mechanisms to reimburse users in case of malfunctions caused by their proprietary oracle modules. Investments appear to be focused on preventing malfunctions through design rather than on ex-post compensation schemes.

**Table 7**. Binary question on compensation for proprietary-oracle malfunction.

| Question | Yes | No |
|---|---|---|
| Does your proprietary oracle have a mechanism to cover in case of a malfunction? | 0% | 100% |

### 4.2.2 Third-Party solutions: Drivers and Barriers

This subsection addresses RQ3 and RQ4 for protocols using third-party oracles. Because selection and multi-sourcing items allowed multiple selections, results are reported as the frequency of selection. As previously shown (Table 5), since multi-sourcing is common, the analysis of third-party solutions begins with the reasons for choosing multiple providers versus a single provider.

As shown in Table 8, the decision to engage multiple third-party oracles is predominantly driven by technical considerations. The majority of respondents indicate that each provider offers specific features critical to the protocol's functioning. A second important driver is multi-chain deployment. When a protocol operates across chains that the main provider does not all support, it is effectively compelled to add additional oracles. Security considerations are also noted, as multiple data sources serve as a backup in case of downtime or reliability issues affecting the primary provider. Because respondents could select multiple reasons, the pattern should be read as a ranking of drivers rather than as exclusive categories.

**Table 8**. Ranking of reasons for choosing multiple third-party oracles

| Respondent Choices | n |
|---|---|
| Feature Complementarity: Each oracle provides unique features that are necessary for our operations. | 17 |
| Chain or Token diversity: Each oracle supports a chain or token that others do not, allowing for broader coverage and functionality. | 16 |
| Redundancy for Security: To ensure continuous service and data integrity in case of failure. | 14 |
| Diversification of Data Sources: To avoid reliance on a single data source | 10 |
| Cost Optimization: Some oracles offer more cost-effective solutions for specific types of data. | 2 |
| Regulatory Compliance: Different oracles may be required to meet different regulatory standards. | 0 |

Table 9 reports the reasons for choosing a specific third-party oracle when multiple options exist for the same network or token. Reputation and security are ranked as the most critical factors, selected by 30 and 25 respondents, respectively. By contrast, innovation and regulation are not considered critical drivers of oracle selection, despite being prominent topics in public discourse around DeFi infrastructure. This suggests that third-party oracle selection is primarily framed as risk and reliability management rather than as experimentation with novel features, while regulatory compliance is rarely an explicit selection driver at the choice stage.

**Table 9**. Ranking of reasons for choosing a specific third-party oracle

| Respondent Choices | n |
|---|---|
| Reputation: “third-party-oracle”’s standing and perceived reliability in the industry. | 30 |
| Security: The robustness of “third-party-oracle”’s security measures | 25 |
| Historical Performance: Track record of the third-party oracle’s past performance and reliability | 12 |
| Network Effects: Advantages associated with the large number of users and applications already integrated with the third-party oracle | 8 |
| Price: Cost-effectiveness of using “third-party-oracle” compared to other oracles. | 6 |
| Community and Support: The strength and responsiveness of “third-party-oracle”’s user and developer community and the quality of technical support. | 6 |
| Unique Features: Specific capabilities or services offered by “third-party-oracle” that are not available with other providers | 2 |
| Innovation Rate: The frequency and impact of “third-party-oracle”’s technological advancements | 0 |
| Regulatory Compliance: “third-party-oracle”’s adherence to legal and regulatory standards. | 0 |

The role of price requires more careful interpretation. Pricing does not emerge as a primary stated selection criterion, which is consistent with the pricing models actually reported by respondents. Most protocols access, in fact, oracle data under “free” arrangements and a substantial share report pay-per-request schemes (Table 10). In this context, “cost” is less about explicit data fees and more about indirect integration, monitoring, and operational burdens associated with relying on an external data layer.

Trust in third-party providers is further examined through a sequence of questions on knowledge of the oracle, awareness of reporters, trust in reporters, and expectations of coverage in case of misreporting. Note that expectations of coverage in case of misreporting are captured as a Yes/No/IDK item (Table 11a), while trust in reporters is captured on a 1–5 scale (Table 11b).

Most respondents (93.55%) report having a clear understanding of how the third-party oracle works. This is unsurprising given that several protocols also operate proprietary oracles and therefore have a high level of technical expertise. Furthermore, integrating a third-party oracle requires developing ad-hoc integration modules, which arguably forces protocols to become familiar with the oracle's functioning.

**Table 10**. The pricing model of your third-party oracle

| Pricing model | Protocols_selecting | % of protocols (among those answering pricing) |
|---|---|---|
| Free | 25 | 80.6 |
| Pay-per-request | 9 | 29 |
| Custom | 3 | 9.7 |
| Subscription | 2 | 6.5 |
| Tiered | 1 | 3.2 |

However, when asked whether they know who the data reporters are, the share of positive responses drops slightly above 60%. Confidence further decreases when respondents are asked whether they consider these reporters trustworthy, as fewer than half (48.4%) say they are sure of their trustworthiness.

**Table 11a**. Awareness and risk allocation

| Question | Yes | % | No | % | I don't know | % |
|---|---|---|---|---|---|---|
| **Awareness** | | | | | | |
| Do you know how your third-party oracle works? | 29 | 93.55 | 2 | 6.45 | - | |
| Do you know who the reporters of the third-party oracle you use are? | 19 | 61.29 | 12 | 38.71 | - | |
| **Risk Allocation** | | | | | | |
| Do you think your third-party oracle will cover in case of misreporting? | 4 | 12.90 | 21 | 67.74 | 6 | 19.35 |

**Table 11b**. Perceived trustworthiness of third-party data reporters.

| Question | Trustworthiness Level | n | % |
|---|---|---|---|
| How trustworthy are the data reporters of the third-party oracle you use? | Very Trustworthy | 15 | 48.4 |
| | Trustworthy | 10 | 32.3 |
| | Neutral | 6 | 19.4 |
| | Non Trustworthy | 0 | 0% |
| | Very Untrustworthy | 0 | 0% |

Expectations regarding compensation in the event of manipulation or misreporting are even more skeptical. Nearly 70% of respondents are convinced that their third-party oracle will not cover losses arising from misreporting, and only a small minority (12.9%) are confident that some remedy mechanisms will be applied. As with proprietary oracles, the focus appears to be on preventing failures rather than on setting up ex post compensation schemes.

The final part of this block explores whether protocols currently relying exclusively on third-party oracles plan to develop a proprietary solution in the future (Table 12a). A large majority (84.21%) do not intend to build an in-house oracle. When asked about potential advantages of doing so, respondents converged again on technical features not currently available from suppliers (Table 12b). Security, reputation, cost, and innovation are also mentioned, but each is mentioned by roughly one-third of respondents at most. This suggests that, for most protocols, the perceived benefits of a proprietary oracle do not outweigh the convenience of outsourcing, except for very specific technical needs. Overall, third-party outsourcing appears to be the default option when functional requirements are met, while the residual demand for in-house development is concentrated in niche technical needs.

**Table 12a**. Intention to build a proprietary oracle among non-proprietary users

| Question | Yes | % | No | % |
|---|---|---|---|---|
| Would you plan to build your own oracle? | 3 | 15.79 | 16 | 84.21 |

**Table 12b**. Ranking of perceived advantages for building a proprietary oracle.

| Respondent choices | n |
|---|---|
| Unique Features: A proprietary oracle would include unique features that a third-party oracle would not guarantee | 12 |
| Security: A proprietary security model would be greater than a third-party one | 6 |
| Innovation Rate: A proprietary oracle would be more upgradable than a third-party one | 5 |
| Reputation: Own reputation is perceived as higher than any third-party oracle's | 5 |
| Price: Cost-effectiveness of using proprietary oracle. | 5 |
| Community and Support: Own protocol community would be more supportive than a third-party oracle community | 3 |
| Regulatory Compliance: A proprietary oracle would guarantee more compliance with regulations | 1 |

### 4.3 Switching costs

This section examines barriers to changing oracle providers. As a first approximation, respondents were asked whether they would switch providers if another offered the same service at a lower price (Table 13). Only 6.25% answered yes, while 93.75% answered no. This indicates that, in the vast majority of cases, lower price alone is insufficient to motivate switching. The pattern is consistent with the presence of switching frictions that go beyond direct monetary cost.

**Table 13**. Binary question on willingness to switch providers

| Question | Yes | % | No | % |
|---|---|---|---|---|
| If offered at a lower price, would you pick another oracle? | 2 | 6.2 | 30 | 93.8 |

Table 14 reports responses to a direct question on how difficult it would be to change or add an oracle provider. Responses are distributed across the scale rather than concentrated at a single point, suggesting that switching difficulty is not perceived uniformly across protocols. To assess whether these perceptions differ systematically by oracle sourcing configuration, we compared switching-difficulty scores across proprietary, third-party, and hybrid users using a Kruskal–Wallis test. The result is not statistically significant ($H(2)=2.91$, $p=0.234$, Table S4b), indicating that reported switching difficulty does not differ systematically across these groups. Likewise, willingness to switch under a cheaper equivalent offer is not associated with oracle structure ($\chi^2(2)=0.19$, $p=0.911$, Table S4a). Together, these results suggest that lock-in is not confined to a specific sourcing architecture, but is instead a broader feature of oracle integration.

Table 14. Perceived difficulty in switching or integrating another oracle.

| Question | Switching Difficulty | n | % |
|---|---|---|---|
| How difficult is it to change or select another oracle? | 1(Not Difficult) | 7 | 21.9 |
| | 2 | 4 | 12.5 |
| | 3(Neutral) | 7 | 21.9 |
| | 4 | 10 | 31.2 |
| | 5(Very Difficult) | 4 | 12.5 |

Because the quantitative results alone do not explain the origin of these frictions, the open-ended responses are particularly informative. Thematic coding of the comments further shows that switching barriers are mainly associated with integration effort, immutability or redeployment constraints (5 occurrences), governance overhead (4 occurrences), and security or audit considerations (2 occurrences).

Some respondents are genuinely satisfied with their provider and simply do not consider alternatives, even if switching would technically be feasible. Others acknowledge that they could add new providers but would need to pass through governance processe**s**, which they interpret as “not particularly difficult” but time-consuming.

More severe constraints emerge among respondents who rated switching as difficult or very difficult. Their comments indicate that, in many cases, changing providers is not realistically feasible. A representative explains:

"*Changing [oracles] requires withdrawing all TVL and going through significant changes; we can add additional ones, but not replace existing.*"

Another representative notes that the protocol's oracle is embedded directly into consensus and block-building, making substitution extremely complex. A further respondent states:

*"(Our protocol) is an immutable system. We cannot alter the smart contracts in any way. The chosen oracles are enshrined in the protocol."*

These statements illustrate how the immutability of smart contracts, the enshrinement of oracles in protocol logic, and governance processes contribute to effective vendor lock-in, even when technical alternatives exist in principle.

4.4 Technology management, acceptance, and performance.

This section addresses RQ6 by examining how protocols use oracle services in practice and how they evaluate the performance of proprietary versus third-party solutions.

Figure 3 summarizes how often protocols require data from their oracle. The pattern suggests an association between request frequency and sourcing choice. Real-time or continuously updated data are more often linked to third-party oracle use, while proprietary solutions appear more common when updates are required less frequently. Overall, lower data intensity is associated with lower dependence on third-party providers.

Figure 3. How often do you require data from your oracle?

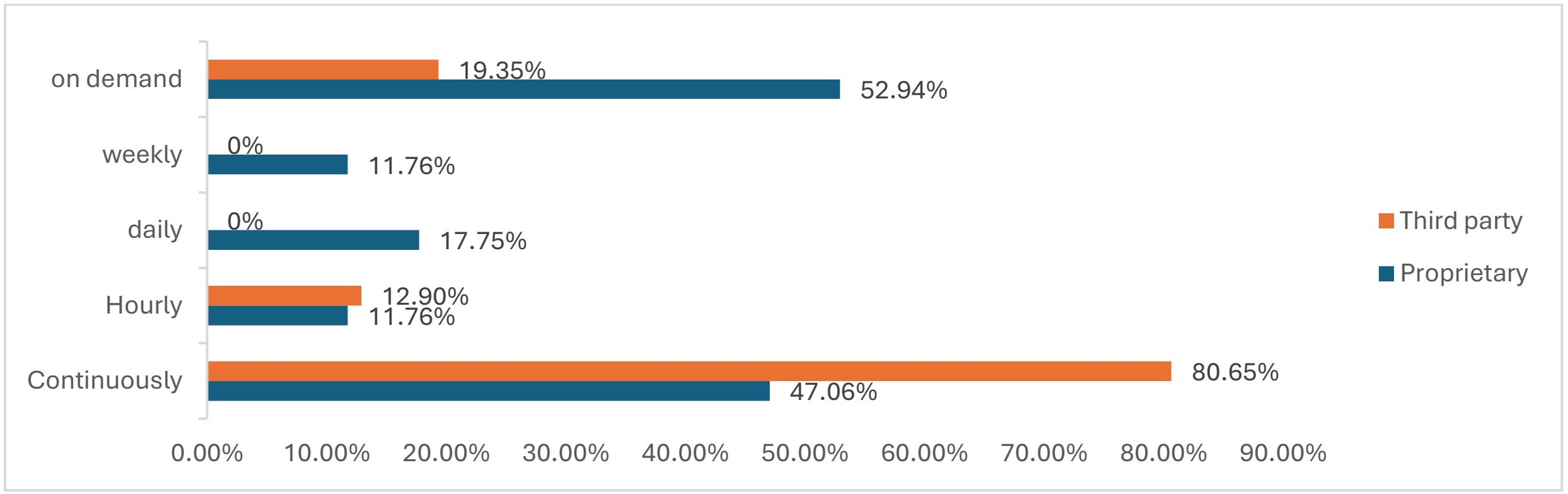

Figure 4 reports the main categories of data requested from oracles and shows how these needs differ between proprietary and third-party solutions. Price feeds are the primary data type for both. Interestingly, RWA data appears more often associated with internal solutions, while inter-chain communication is more frequently delegated to third-party oracles. When operating across multiple chains, protocols appear to prefer deploying multi-chain versions of their core product rather than relying on third-party cross-chain bridges or messaging systems. This suggests that, where a viable alternative exists, protocols tend to minimize dependence on third-party intermediaries for cross-chain operations.

Figure 4. What data do you require from your oracle?

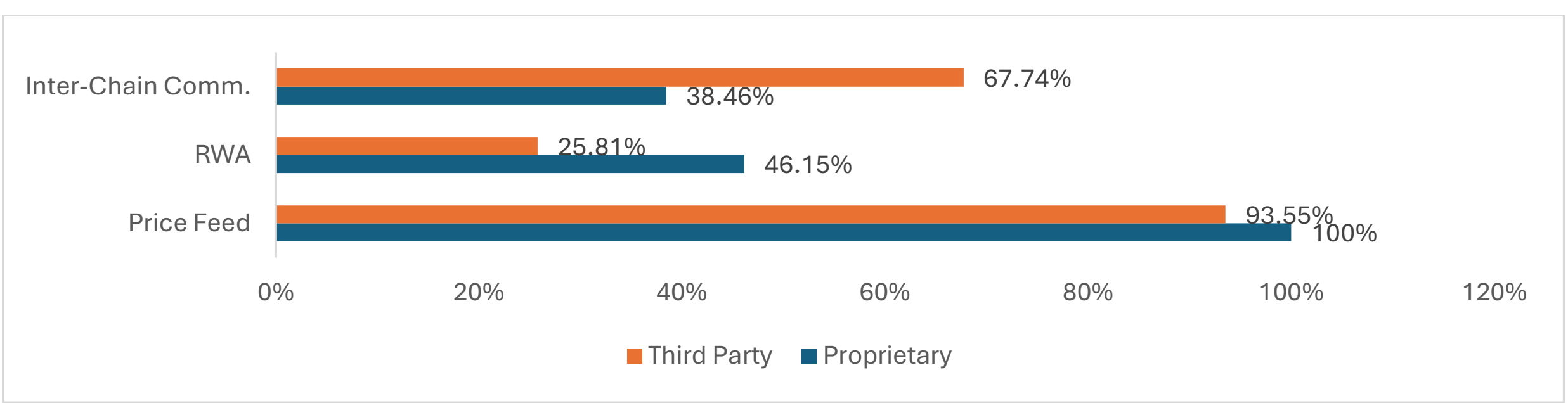

Perceptions of service quality are captured through 1-to-5 evaluations of timeliness, completeness, correctness, and overall satisfaction (Table 15). For both proprietary and third-party oracles, average scores are above 4, indicating generally high satisfaction with oracle performance. Third-party solutions nevertheless receive descriptively higher average scores on all four indicators. To summarize these dimensions jointly, we computed a composite Satisfaction Index as the mean of the four items. The internal consistency of this block is moderate for both third-party and proprietary evaluations (Cronbach's $\alpha = 0.598$ and 0.665, respectively; Table S5a). Among hybrid protocols that evaluated both solutions, a paired Wilcoxon signed-rank test does not show a statistically significant difference between third-party and proprietary satisfaction indices ($W = 33$, $p = 1.000$, $r = 0.136$; Table S5b). Thus, although third-party solutions are descriptively rated more positively, the paired evidence from hybrid protocols does not indicate a statistically robust difference in overall satisfaction.

**Table 15**. Perception of data quality and overall satisfaction

| **Items** | **Proprietary** | **Third Party** |
| --- | --- | --- |
| | **Mean values (1-5)** | |
| Timeliness | 4.08 | 4.32 |
| Completeness | 4.23 | 4.39 |
| Correctness | 4.15 | 4.23 |
| Overall Satisfaction | 4.15 | 4.55 |
| **Satisfaction index** | | |
| Mean of timeliness, completeness, correctness, and overall satisfaction | 4.15 | 4.37 |

The final part of the analysis compares the perceived characteristics of proprietary and third-party oracles, as well as their relative position compared to other available solutions in the market. Table 16 reports two complementary assessments: mean 1–5 evaluations of each attribute and relative comparisons against other available solutions (“less”, “same”, “more”), which are analyzed descriptively through category distributions. Reputation and community support are measured only for third-party oracles, as these attributes are externally oriented by definition.

Third-party oracles receive very high average scores for reputation (4.65) and security (4.55), slightly outperforming proprietary solutions on the latter (4.23). Perceptions of decentralization and innovation and adherence to legal standards are somewhat lower but remain positive and broadly similar across proprietary and third-party solutions.

Regarding cost, both proprietary and third-party solutions are rated as not very expensive. Proprietary solutions are perceived as slightly less costly, despite the fact that many third-party feeds are accessed under free or pay-per-request models. This likely reflects the hidden integration and optimization burdens associated with connecting to external providers. The clearest difference concerns ease of use. Third-party oracles are perceived as easier to use than proprietary solutions, suggesting that outsourcing the data-gathering layer shifts part of the burden of monitoring, source selection, and security management to the external provider, thereby reducing operational complexity for the protocol team.

**Table 16**. Characteristics comparison between proprietary (P) and third-party (T) oracle.

| P/T* | Attribute | Evaluation (1-5) Average value | Compared with others (%) | | |
|---|---|---|---|---|---|
| | | | Less | Same | More |
| T | Reputation | 4.65 | 6.5 | 9.7 | 83.9 |
| T | Community Support | 4.19 | 0 | 64.5 | 35.5 |
| T | Decentralization | 3.90 | 3.2 | 51.6 | 45.2 |
| P | Decentralization | 4 | 0 | 69.2 | 30.8 |
| T | Security | 4.55 | 0 | 25.8 | 74.2 |
| P | Security | 4.23 | 0 | 53.8 | 46.2 |
| T | Expensiveness | 2.65 | 6.5 | 74.2 | 19.4 |
| P | Expensiveness | 2.38 | 15.4 | 53.8 | 30.8 |
| T | Regulatory Adherence | 3.94 | 0 | 74.2 | 25.8 |
| P | Regulatory Adherence | 3.77 | 0 | 84.6 | 15.4 |
| T | Innovation | 3.71 | 6.5 | 61.3 | 32.3 |
| P | Innovation | 3.62 | 7.7 | 61.5 | 30.8 |
| T | Ease of Use | 3.65 | 0 | 74.2 | 25.8 |
| P | Ease of Use | 3.00 | 15.4 | 84.6 | 0 |

*P=Proprietary, T=Third-party

Finally, the relative-comparison responses suggest that respondents are generally cautious in rating proprietary solutions as superior to market alternatives. Even when absolute evaluations of proprietary oracles are positive, they are more often described as “about the same” than as clearly “better than others”. This indicates a conservative comparative stance and may reflect the tendency of respondents to avoid benchmarking internal solutions against well-established external providers.

### 4.5. Exploratory pairwise associations among evaluation variables.

To complement the descriptive and comparative analyses reported above, Table 17 presents a Spearman rank-order correlation matrix for the 31 protocols using third-party oracle solutions. The matrix covers ordinal evaluation variables (switching difficulty, trust in reporters, satisfaction index, and attribute ratings) as well as two derived count variables (number of oracle providers and number of supported chains). Three associations reach conventional significance ($p < 0.05$). First, perceived security and perceived decentralization of the third-party oracle are strongly and positively correlated ($\rho = 0.611$, $p < 0.001$), suggesting that respondents perceive these attributes as a coherent bundle rather than as independent dimensions. Second, the satisfaction index is negatively associated with perceived innovation ($\rho = -0.472$, $p = 0.007$), indicating that more satisfied protocols rate their third-party oracle as less innovative. Third, perceived expensiveness is negatively correlated with perceived reputation ($\rho = -0.458$, $p = 0.010$), consistent with the observation that the most reputable provider operates under free or low-cost pricing models. Additional marginal associations ($p < 0.10$) include a positive link between decentralization and innovation ($\rho = 0.392$, $p = 0.029$), between trust in reporters and community support ($\rho = 0.337$, $p = 0.064$), and negative associations between the number of oracle providers used and both reputation ($\rho = -0.317$, $p = 0.083$) and community support ($\rho = -0.335$, $p = 0.065$) ratings. Notably, switching difficulty does not correlate significantly with any other variable in the matrix.

**Table 17.** Spearman rank-order correlation matrix (third-party oracle evaluations, n = 31).

| Variable | (1) | (2) | (3) | (4) | (5) | (6) | (7) | (8) | (9) | (10) | (11) | (12) | (13) |
|---|---|---|---|---|---|---|---|---|---|---|---|---|---|
| (1) Switching Difficulty | 1.000 | | | | | | | | | | | | |
| (2) Trust in Reporters | -0.264 | 1.000 | | | | | | | | | | | |
| (3) Satisfaction Index (TP) | -0.266 | 0.216 | 1.000 | | | | | | | | | | |
| (4) Security (TP) | 0.043 | 0.129 | 0.031 | 1.000 | | | | | | | | | |
| (5) Decentralization (TP) | 0.091 | -0.071 | -0.096 | 0.611*** | 1.000 | | | | | | | | |
| (6) Expensiveness (TP) | -0.196 | -0.114 | -0.220 | 0.005 | 0.220 | 1.000 | | | | | | | |

| Variable | (1) | (2) | (3) | (4) | (5) | (6) | (7) | (8) | (9) | (10) | (11) | (12) | (13) |
|---|---|---|---|---|---|---|---|---|---|---|---|---|---|
| (7) Regulatory Adherence (TP) | -0.046 | 0.096 | 0.065 | -0.263 | -0.236 | -0.229 | 1.000 | | | | | | |
| (8) Innovation (TP) | 0.139 | -0.211 | -0.472** | 0.297 | 0.392* | 0.155 | 0.207 | 1.000 | | | | | |
| (9) Ease of Use (TP) | -0.055 | -0.141 | 0.141 | 0.326† | 0.146 | 0.020 | -0.183 | 0.052 | 1.000 | | | | |
| (10) Reputation (TP) | 0.039 | 0.201 | 0.148 | -0.069 | 0.070 | -0.458** | -0.186 | -0.066 | 0.067 | 1.000 | | | |
| (11) Community Support (TP) | -0.079 | 0.337† | 0.326† | -0.009 | 0.090 | -0.036 | 0.033 | -0.241 | -0.093 | 0.245 | 1.000 | | |
| (12) N. of Providers | 0.101 | -0.247 | -0.059 | 0.102 | 0.128 | 0.008 | -0.254 | 0.058 | 0.075 | -0.317† | -0.335† | 1.000 | |
| (13) N. of Chains | 0.142 | -0.130 | 0.042 | -0.097 | 0.198 | -0.196 | -0.016 | 0.028 | -0.085 | 0.166 | -0.165 | 0.344† | 1.000 |

Lower triangle reports Spearman's ρ. The sample includes 31 protocols using third-party oracle solutions (19 third-party only + 12 hybrid); the one proprietary-only protocol is excluded. TP = Third-party evaluations. N. of Providers and N. of Chains are derived count variables.
*** $p < 0.001$, ** $p < 0.01$, * $p < 0.05$, † $p < 0.10$.

An analogous matrix was computed for the 13 protocols evaluating proprietary oracle solutions (Table S6). Given the smaller sample, only strong associations reach conventional significance. Satisfaction with proprietary oracles is strongly correlated with both perceived security ($\rho = 0.808$, $p < 0.001$) and decentralization ($\rho = 0.826$, $p < 0.001$), suggesting that for protocols that build in-house, satisfaction is driven by the achievement of the security and decentralization properties they deliberately pursued. Notably, the association between satisfaction and innovation is positive and marginally significant ($\rho = 0.528$, $p = 0.064$), in contrast with the negative association observed in the third-party matrix. These patterns are further interpreted in the Discussion. The following section discusses these results and positions the research within the supplier selection literature.

## 5. Discussion

This section interprets the findings in light of the literature on supplier selection, IT outsourcing, and switching costs in digital infrastructure markets. Across the six research themes, the results suggest that oracle adoption in DeFi is not only a technical design choice, but also a sourcing and governance decision shaped by functional fit, provider reputation, architectural embedding, and reversibility constraints. The discussion is organized by theme to preserve continuity with the empirical analysis and to clarify how the findings extend existing research on digital-service sourcing and dependence.

### 5.1 Supplier selection criteria in the oracle context

Classical supplier-selection research considers purchasing decisions as multi-criteria processes involving trade-offs among quality, reliability, price, technical capability, and other organizational concerns (Dickson, 1966; Weber, Current and Benton, 1991). The present findings suggest that this logic remains valid in the oracle market, but that the relative importance of criteria is reshaped by the technological and organizational characteristics of DeFi. In particular, reputation and security emerge as the most important reasons for choosing third-party oracle providers, whereas price appears to play a comparatively limited role.

This pattern is consistent with the fact that oracle services are often perceived not as ordinary inputs, but as mission-critical infrastructure. In such a setting, the cost of a provider is less important than the potential consequences of malfunction, manipulation, or operational unreliability. The weak role of price may also reflect the prevailing pricing models in the sector, where some feeds are accessed for free and others under pay-per-request structures that are not perceived as especially burdensome. As a result, the supplier-selection problem shifts away from a conventional cost–quality trade-off and toward the management of security, trust, and ecosystem compatibility. Similar dynamics have been observed in cloud-service selection, where reliability, compliance, and security often dominate purely economic criteria (Lang, Wiesche and Krcmar, 2016). The correlation analysis reinforces this interpretation. Respondents who rate third-party oracles highly on security also rate them highly on decentralization, implying these attributes are evaluated jointly rather than in isolation. This aligns with evidence from cloud-provider evaluation, where trust-related attributes cluster in user perceptions (Luna *et al.*, 2017). Moreover, the most reputable providers are also perceived as the least expensive, effectively neutralizing cost as a differentiator and explaining its low importance in selection. This mirrors network-effect markets (Katz and Shapiro, 1994), where dominant providers' scale advantages compress costs and limit price-based competition. Taken together, these results suggest that oracle selection reproduces the multi-criteria logic of supplier choice, but in a way that reweights traditional dimensions such as quality and cost toward security, reputation, interoperability, and mission-critical reliability.

### 5.2 Make-or-buy decisions and path dependence in oracle sourcing

The second set of findings concerns the make-or-buy decision for oracle functionality. IT outsourcing research has long emphasized that decisions to internalize or externalize ICT services depend on transaction costs, internal capabilities, asset specificity, and risk (Opara-Martins, Sahandi and Tian, 2015).

Our data show that protocols primarily build proprietary oracles to obtain technical features that third-party providers do not (or did not) offer, such as specific time-weighted or volume-weighted price mechanisms, support for illiquid tokens, or custom security checks, rather than to reduce fees or increase independence from suppliers. This finding resonates with resource-based views of outsourcing that emphasize the importance of unique technical capabilities and service specificity, as when off-the-shelf solutions do not meet the buyer's specific needs, internal development becomes more attractive despite higher costs (Hassanzadeh and Cheng, 2016).

However, our qualitative evidence also reveals a strong element of path dependence. Early protocols built their own oracles because third-party options were either absent or immature. Later improvements in external services did not necessarily trigger migration away from the in-house solution. Once a proprietary oracle has been heavily audited, parameterized, and integrated into governance processes, its replacement is perceived as both risky and costly, even when functionally equivalent services become available on the market. This dynamic reflects findings from legacy system and ERP migration studies, where technological and organizational inertia constrain the adoption of superior solutions (Davis, 2015; Irani *et al.*, 2023). The immutability of smart contracts and the enshrinement of oracle logic into protocol code further amplify this path dependence compared to traditional IT settings.

Interestingly, perceptions of difficulty in building and maintaining proprietary oracles are not homogeneous. Some respondents describe a substantial ongoing burden, continuous tuning, audits, and governance overhead, while others depict the oracle as a relatively simple data pipe, shifting the main responsibility for robustness to protocol design. This divergence suggests that the boundary between "oracle work" and "protocol work" is itself contested. For some teams, security and robustness are primarily engineered into the oracle subsystem, while for others, they are addressed at the contract design and liquidation logic levels. From a theoretical perspective, this highlights the need to better understand how responsibilities for reliability and risk management are partitioned across layers of digital infrastructure. These findings suggest that future research should broaden the theoretical discussion of oracle architecture and implementation patterns initiated by Pasdar et al. (2023)

Overall, our findings extend make-or-buy discussions by showing how technical specificity, historical timing, and protocol immutability combine to shape oracle sourcing decisions in ways that standard cost–capability models only partially capture.

### 5.3 Multi-sourcing, lock-in, and the paradox of flexibility

A third contribution concerns multi-sourcing and vendor lock-in. Supplier selection and multi-criteria decision-making studies often treat multi-sourcing as a strategy to mitigate risk, balance performance across suppliers, and improve bargaining power (de Boer, Labro and Morlacchi, 2001). In cloud computing, multi-cloud strategies are explicitly recommended as a way to reduce dependence on a single provider and avoid lock-in. At the same time, an extensive literature documents the persistence of vendor lock-in due to proprietary standards, data formats, and integration costs, with Opara-Martins et al. (2016) identifying lock-in as a major barrier to cloud migration.

In our sample, multi-sourcing is instead almost a standard for oracle supply. Many protocols combine multiple third-party oracles, and a sizeable fraction add a proprietary module on top. At face value, this appears consistent with risk-mitigation strategies in IT sourcing. Yet the detailed findings indicate different underlying motives. First, multiple providers are often used not to create symmetric redundancy, but to cover non-overlapping functional or network niches (different chains, different asset types, specialized features). This limits the substitutability between suppliers and may actually increase overall dependence on the combined oracle "stack".

Second, quantitative responses on switching intentions and difficulty reveal a sort of paradox of flexibility. A large majority of respondents would not switch to another provider even if it offered the same service at a lower price, and many describe switching as difficult or very difficult. Qualitative comments clarify that, for some protocols, changing an oracle would require withdrawing TVL, modifying core contracts, or even rewriting components embedded in the consensus or block-building logic. In several cases, respondents explicitly state that the system is immutable and that the current oracle configuration is "enshrined" in the protocol. At the same time, the inferential checks reported in the findings show that neither switching intention nor perceived switching difficulty varies significantly across proprietary, third-party, and hybrid sourcing configurations, suggesting that lock-in is a broader feature of oracle integration rather than a problem confined to one specific architecture. The correlation analysis supports this conclusion. Perceived switching difficulty is unrelated to satisfaction, trust, security, or any other evaluation dimension, indicating that protocols are not locked in because they are dissatisfied, but because of architectural constraints. This contrasts with traditional switching-cost models, where dissatisfaction typically drives switching intentions (Bansal, 2005), and instead aligns with structural lock-in in cloud computing, where exit barriers depend on integration depth rather than service quality (Opara-Martins, Sahandi and Tian, 2016).

Even when multiple providers exist, integration costs, proprietary interfaces, and architectural constraints can make switching prohibitively costly. In the oracle context, the situation is further compounded by blockchain immutability, as once an oracle is deeply embedded in protocol logic, the theoretical modularity of smart contracts does not translate into practical substitutability. In this setting, therefore, multi-sourcing does not necessarily guarantee exit options. It results instead in layered dependence, where a dominant provider remains indispensable while smaller providers are added for specific purposes.

From a theoretical standpoint, this suggests that models of supplier selection and outsourcing in digital infrastructures must account not only for the number of suppliers but also for the architecture of integration, that is, how deeply a supplier's service is embedded into the client's core logic, and whether that embedding is reversible.

**5.4 Risk allocation, trust, and perceived legal compliance**

Our findings also help clarify how risk and responsibility are allocated between protocols and oracle providers. IS outsourcing research has traditionally emphasized the role of trust, contract completeness, and service-level agreements in defining who bears the consequences of failures (Hassanzadeh and Cheng, 2016). Cloud studies further highlight uncertainty about liability in the event of breaches or outages as a dimension of perceived lock-in and adoption risk (Opara-Martins, Sahandi and Tian, 2016).

In our survey, both proprietary and third-party oracles are evaluated very positively on ex-ante performance (timeliness, completeness, correctness). At the same time, almost no proprietary protocol has mechanisms to reimburse users in case of oracle failures, and a large majority of clients do not expect third-party providers to cover losses arising from misreporting, regardless of whether the oracle protocol formally advertises insurance or compensation schemes. In practice, they behave as if residual risk remains with the protocol. In other words, protocols are highly satisfied with how oracles perform in normal conditions but assume that extreme failures will be borne by the protocol and its users, not by the oracle supplier.

This pattern suggests a distinctive form of risk internalization. Rather than relying on ex-post compensation or formal liability, protocols invest in ex-ante robustness, through design choices such as conservative price feeds, TWAP windows, fallback mechanisms, and governance safeguards, and accept residual risk as part of their own operational exposure. Conceptually, this differs from many traditional outsourcing settings, where detailed contracts specify penalties and remedies. It aligns more closely with arrangements in which a critical infrastructure is used "as is" and clients protect themselves primarily via redundancy and resilience measures.

A second interesting asymmetry concerns knowledge and trust. Most clients report understanding how the third-party oracle works, yet fewer than half feel certain about the trustworthiness of the reporters feeding it. This indicates a mechanism-level trust (in the oracle design) combined with agent-level opacity (about who actually reports data). It aligns with concerns in the cloud and platform governance literature that users may trust the technical architecture while remaining uncertain about the actors behind it (Huang and Nicol, 2013; Lynn *et al.*, 2021).

A further pattern qualifies the relationship between satisfaction and innovation across sourcing configurations. Among protocols relying on third-party oracles, higher satisfaction is associated with lower perceived innovation, suggesting a preference for stability over novelty, consistent with supplier-selection literature on mission-critical inputs (Dickson, 1966; Weber, Current and Benton, 1991). In contrast, for protocols using proprietary oracles, satisfaction increases with perceived innovation, reflecting the intentional development of features unavailable from external providers.

This opposing pattern indicates that innovation plays a different role depending on sourcing; therefore, it is a source of value when internally developed, but a potential source of instability when externally imposed. To our knowledge, this distinction has not been documented in the IT outsourcing literature, where innovation is typically treated as uniformly beneficial.

Finally, there is a discrepancy between the low salience of legal and regulatory compliance in selection decisions and the high ex post perceived legal adherence of third-party oracles. Legal aspects are rarely cited as primary reasons for choosing a provider, yet third-party oracles are rated as more compliant with legal standards than proprietary solutions, although the gap is moderate rather than sharp. Given the current regulatory opacity surrounding oracles in most jurisdictions, this likely reflects assumptive trust, the belief that established providers "must have" addressed legal issues more thoroughly, rather than a detailed assessment of their actual legal position. This suggests a potential decoupling between formal compliance and perceived legitimacy in the adoption of decentralized infrastructure. To consolidate the interpretation of results, Table 18 summarizes the main findings for each research theme and the corresponding contributions to current literature.

**Table 18**. Summary of findings and implications by research theme

| Research Theme | Main Findings | Implications for academic literature |
|---|---|---|
| Protocol and sourcing context | -DeFi oracle client side is dominated by Lending/asset management protocols operating on EVM chains<br><br>-Oracle architecture is hybrid and multisource<br><br>-A single dominant provider appears in almost all configurations, with niche competitors added for specific assets or chains | -Reflecting ICT literature, a small set of critical suppliers serves almost the whole market.<br><br>-The supplier landscape itself is part of understanding risks and dependencies. |
| Make-or-buy of the oracle module | -Proprietary oracles are built mainly to implement very specific technical features (illiquid assets, custom price | -Technical specificity, historical timing, and smart-contract immutability jointly drive internalization decisions. |
| | logic, conservative TWAP/VWAP design) or because no adequate 3rd-party solution existed at launch.<br><br>-Oracle maintenance is costly (tuning, audits, governance), yet some teams see the oracle as a “simple data pipe” and shift robustness to protocol design.<br><br>-Once deployed and audited, internal solutions are rarely abandoned, creating strong path dependence. | -Emerges a new scenario of “legacy IT” in which on-chain immutability and audit costs make later migration to superior external services difficult, even as the market evolves. |
| Drivers of third-party selection and multi-sourcing | -Third-party selection is driven primarily by reputation, security, and technical/ecosystem fit, while explicit pricing, regulation, and innovation rank lower.<br><br>-Multi-sourcing is common but mostly used to cover non-overlapping needs (different chains, assets, or features) rather than to create symmetric redundancy.<br><br>-The dominant provider remains almost always present, even in multi-provider setups. | -The multi-criteria nature of supplier choice is confirmed, but in Web3, classical criteria such as quality, delivery, and price are re-weighted towards security, reputation, and ecosystem integration.<br><br>-In our sample, multi-sourcing operates primarily as a complement to a dominant supplier rather than as a symmetric redundancy strategy and thus offers limited risk diversification. |
| Trust, knowledge, and risk allocation in third-party relationships | -Most respondents understand how the oracle mechanism works, but a smaller share knows who the reporters are, and even fewer are certain of their trustworthiness.<br><br>-A large majority do not expect compensation from providers in case of misreporting or manipulation.<br><br>-Protocols invest heavily in ex-ante robustness (design, parameters, redundancy) rather than ex-post coverage risk is largely internalized by the protocol and its users. | -Extends IS-outsourcing and cloud-trust literature by documenting a pattern of mechanism-level trust with actor-level opacity: users trust the architecture but have limited visibility on the agents behind it.<br><br>-Challenges the assumptions that outsourcing critical services automatically transfers liability. Instead, risk remains primarily on the client side, with suppliers offering high ex-ante performance but weak ex-post remedies. |
| Switching costs and lock-in | -Almost all respondents would not switch providers even for an equivalent, cheaper service.<br><br>-Self-reported switching difficulty is mixed, but qualitative evidence shows that changing oracles would require withdrawing TVL, modifying immutable contracts, or altering consensus/block-building logic.<br><br>-Multi-sourcing often adds providers, but does not make the dominant one replaceable. | -In Web3, the concept of lock-in and path dependencies is amplified due to smart-contract immutability and deep architectural embedding, even where alternatives exist.<br><br>-Suggests that models of vendor lock-in in digital infrastructures must consider not just the competitiveness of suppliers but the depth and reversibility of integration into core protocol logic. |
| Oracle usage, dependence, and performance | -Third-party oracles are used more for real-time and cross-chain data, while proprietary solutions are more common for lower-frequency updates and some RWA data.<br><br>-Protocols often avoid relying on 3rd-party bridges when they can instead deploy multi-chain versions of their product.<br><br>-Both internal and external oracles score highly on timeliness, completeness, and correctness, but third-party solutions | -Extends vendor-evaluation and outsourcing-satisfaction literature to Web3 by showing when and why protocols prefer outsourcing versus internal development.<br><br>-Highlights a tension between perceived superior performance of third-party oracles and structural dependence on a |

|  | receive higher satisfaction scores, are seen as more legally compliant and easier to use, and are not perceived as significantly more expensive. | concentrated set of providers, reinforcing the importance of studying resilience, standardization, and potential systemic risk in oracle markets. |
|---|---|---|

## 6. Conclusion

This study contributes to the emerging literature on blockchain oracles by examining oracle supplier selection from the perspective of DeFi protocols. Drawing on survey responses from 32 protocols representing 55.4% of the TVL within the DeFined eligible population, it shows that oracle adoption is shaped not only by technical considerations, but also by sourcing logic, provider reputation, architectural embedding, and switching constraints. By conceptualizing oracle providers as specialized ICT suppliers, the paper extends the analysis of oracles beyond design and performance, toward the organizational conditions under which these services are selected and governed.

The findings indicate that protocols generally rely on third-party oracle solutions when external providers satisfy their functional requirements, while proprietary development persists mainly in cases of technical specificity, unsupported assets, or historical timing. Multi-sourcing is common, but often serves complementary rather than substitutive purposes. Once oracle logic becomes embedded into protocol architecture, smart-contract immutability and redesign costs can make switching difficult even when alternatives exist. The study therefore shows that oracle sourcing in DeFi is not merely a matter of technical performance, but also of dependence, reversibility, and infrastructural control.

The study has several limitations. First, the evidence is exploratory and based on a relatively small number of key informants, which limits statistical generalizability across the broader Web3 ecosystem. Second, given market volatility and the need to collect data from active and operating protocols, the sample is biased toward protocols with relatively large TVL, and therefore does not reflect the managerial choices of smaller protocols with substantially lower TVL. Third, the rapid evolution of smart contracts and protocol architectures means that sourcing configurations may change over time, even though immutability can simultaneously slow practical switching. Fourth, the pairwise correlation analysis is bivariate and exploratory; with 31 observations for third-party evaluations and 13 for proprietary evaluations, the analysis can detect only relatively strong associations and cannot control for confounding variables. The patterns identified, including the reversed role of innovation across sourcing configurations, should therefore be treated as indicative directions warranting confirmation in larger and potentially longitudinal samples. Leveraging topics emerged in this study, further research could investigate how protocol characteristics influence sourcing strategies, how concentration evolves in oracle markets, and how governance or regulatory mechanisms may alter the allocation of responsibility when oracle failures occur.

**Appendix A**. Supplementary tables and inferential tests.

This appendix reports supplementary information that does not change the findings but provides additional detail on the sample and analytical outputs. Table S1 summarizes the sample's TVL distribution as measured at the start of data collection (07/02/2025). TVL is strongly right-skewed as the mean is about USD 1.0B, while the median is USD 137M, indicating that a small number of very large protocols drive the average.

**Table S1**. TVL (USD) distribution of the sample

| Metric | Value |
|---|---|
| N-observation | 32 |
| Mean TVL | 1038422500 |
| Median TVL | 137000000 |
| Min TVL | 14500000 |
| Max TVL | 20290000000 |

Table S2 instead reports the distribution of providers by protocols. The data shows an average of 2.4 oracle providers (median = 2, Interquartile Range ≈ 1.75–3; range 0–5), suggesting that multi-provider sourcing is common rather than exceptional.

**Table S2.** Summary of Oracle providers per protocol.

| Stat | value |
|---|---|
| count | 32 |
| mean | 2.40625 |
| std | 1.240691 |
| min | 0 |
| 25% | 1.75 |
| 50% | 2 |
| 75% | 3 |
| max | 5 |

The complete coding result, which classifies and groups keywords in recurrent themes, is observable in Table S3.

**Table S3**. Coding mapping summary.

| Q_id | Question theme | Assigned Theme | Count |
|---|---|---|---|
| Q25 | Third-Party_Main_Barrier | Price accuracy and liquidation risk | 3 |
| Q25 | Third-Party_Main_Barrier | Centralization and admin control | 1 |
| Q25 | Third-Party_Main_Barrier | Cost | 1 |
| Q25 | Third-Party_Main_Barrier | Dependency on provider | 1 |
| Q25 | Third-Party_Main_Barrier | Low barrier / strong support | 1 |
| Q46 | Switching_Cost | Architectural lock-in / immutability | 5 |
| Q46 | Switching_Cost | Governance and organizational approval | 4 |
| Q46 | Switching_Cost | Technical integration and code changes | 4 |
| Q46 | Switching_Cost | Security and risk aversion | 2 |
| Q46 | Switching_Cost | Ecosystem dependence | 1 |
| Q46 | Switching_Cost | Status quo / no need to switch | 1 |
| Q7 | Proprietary_Main_Advantage | Asset coverage | 6 |
| Q7 | Proprietary_Main_Advantage | Custom pricing methodology | 4 |
| Q7 | Proprietary_Main_Advantage | Security and manipulation resistance | 4 |
| Q7 | Proprietary_Main_Advantage | Flexibility and control | 1 |
| Q8 | Proprietary_Main_Challenge | Parameter tuning and methodology | 4 |
| Q8 | Proprietary_Main_Challenge | Reliability and security robustness | 3 |
| Q8 | Proprietary_Main_Challenge | Technical and infrastructure complexity | 3 |
| Q8 | Proprietary_Main_Challenge | Development and audit burden | 2 |

The following tables report the inferential tests related to switching costs and lock-in. In particular, Table S4a presents the contingency table and chi-square test for willingness to switch to a cheaper equivalent oracle provider across proprietary, third-party, and hybrid protocols, as summarized in Table 13 of the main text. This test is used to verify whether switching intention is systematically associated with oracle sourcing configuration. Table S4b reports the Kruskal–Wallis test for differences in perceived switching difficulty across the same sourcing configurations, as reported in Table 14 of the main text. This test is used to assess whether perceived switching difficulty varies significantly across protocols with different oracle structures.

**Table S4a**. Inferential test for switching intention under a cheaper equivalent offer

| Outcome | Predictor | Test | N | chi2 | df | p_value | Effect_size | Effect_value |
|---|---|---|---|---|---|---|---|---|
| Switch to a cheaper equivalent | Oracle structure | Chi-square test of independence | 32 | 0.187 | 2 | 0.911 | Cramer's V | 0.076 |

**Table S4b**. Inferential test for perceived switching difficulty across oracle structures.

| Outcome | Grouping_variable | Test | N | H | df | p_value | Effect_size | Effect_value |
|---|---|---|---|---|---|---|---|---|
| Switching Difficulty (1-5) | Oracle Structure | Kruskal-Wallis | 32 | 2.906 | 2 | 0.234 | Epsilon-squared | 0.031 |

The following tables report supplementary statistical checks related to the service-satisfaction analysis discussed in Section 4.4. Because the four satisfaction items (timeliness, completeness, correctness, and overall satisfaction) are also used as a composite block, Table S5a reports Cronbach's alpha to assess their internal consistency and to verify whether they can be treated jointly as a Satisfaction Index. In turn, because hybrid protocols evaluated both proprietary and third-party solutions, Table S5b reports a Wilcoxon signed-rank test to assess whether the paired satisfaction evaluations differ systematically within the same protocols. In particular, we took the 12 hybrid protocols, computed for each one a third-party satisfaction index and a proprietary satisfaction index, calculated the difference between the two, ranked those differences by size, and used the Wilcoxon signed-rank test to check whether one provider type was systematically rated more favorably than the other. The result showed no significant paired difference. Together, these tests complement the descriptive evidence reported in the main text by checking both the coherence of the satisfaction construct and the robustness of the observed differences between provider types.

**Table S5a**. Internal consistency of the satisfaction block.

| ProviderType | Complete answers per case | Cronbach_alpha |
|---|---|---|
| Third-party | 31 | 0.598 |
| Proprietary | 13 | 0.665 |

**Table S5b**. Paired comparison of satisfaction across provider types.

| Outcome | Test | N_pairs | W | p_value | Effect_size | Effect_value |
|---|---|---|---|---|---|---|
| Paired comparison (Third-Party, proprietary, Hybrid) | Wilcoxon signed-rank | 12 | 33 | 1 | r | 0.136 |

Finally, Table S6 reports the Spearman rank-order correlation matrix for the 13 protocols that evaluated their proprietary oracle solution. The matrix includes proprietary attribute ratings, the satisfaction index, switching difficulty, and number of supported chains. Trust in reporters and number of providers are excluded as they pertain exclusively to third-party evaluations. The most notable difference relative to the third-party matrix (Table 17) is the reversed sign on the satisfaction/innovation association, which is discussed in Section 5.4. Given the small sample size, only strong associations reach conventional significance, and results should be interpreted as exploratory.

**Table S6.** Spearman rank-order correlation matrix (proprietary oracle evaluations, n = 13).

| Variable | (1) | (2) | (3) | (4) | (5) | (6) | (7) | (8) | (9) |
|---|---|---|---|---|---|---|---|---|---|
| (1) Switching Difficulty | 1.000 | | | | | | | | |
| (2) Satisfaction Index (P) | 0.047 | 1.000 | | | | | | | |
| (3) Security (P) | 0.064 | 0.808*** | 1.000 | | | | | | |
| (4) Decentralization (P) | 0.151 | 0.826*** | 0.541† | 1.000 | | | | | |
| (5) Expensiveness (P) | 0.510† | 0.049 | 0.279 | -0.157 | 1.000 | | | | |
| (6) Regulatory Adherence (P) | 0.286 | 0.093 | 0.083 | 0.271 | 0.164 | 1.000 | | | |
| (7) Innovation (P) | 0.286 | 0.528† | 0.396 | 0.459 | 0.046 | 0.027 | 1.000 | | |
| (8) Ease of Use (P) | -0.252 | 0.437 | 0.125 | 0.486† | -0.214 | -0.248 | 0.397 | 1.000 | |
| (9) N. of Chains | -0.036 | -0.242 | 0.097 | -0.326 | 0.054 | -0.402 | 0.355 | -0.029 | 1.000 |

Lower triangle reports Spearman's $\rho$ ($n = 13$ for all pairs). The sample includes 13 protocols that evaluated their proprietary oracle solution (12 hybrid + 1 proprietary-only). P = Proprietary evaluations.
*** $p < 0.001$, ** $p < 0.01$, * $p < 0.05$, † $p < 0.10$.